\documentclass[12pt]{iopart}

\usepackage{iopams}
\usepackage{setstack}
\usepackage{enumerate}

\newcommand{\mb}[1]{\mathbf{#1}} 
\newcommand{\R}{\mathbb{R}}      
\newcommand{\C}{\mathbb{C}}      
\newcommand{\Int}{\mathrm{int}}  
\newtheorem{Theorem}{Theorem}    
\newtheorem{Lemma}{Lemma}        

\begin{document}
\title[The principle of equivalence and projective structure in space-times]{The principle of equivalence and projective structure in space-times}
\author{G S Hall$^1$ and D P Lonie$^2$ }
\address{$^1$Department of Mathematical Sciences, University of Aberdeen,
Meston Building, Aberdeen, AB24 3UE, Scotland, U.K.}
\address{$^2$108e Anderson Drive, Aberdeen, AB15 6BW, Scotland, U.K.}
\eads{\mailto{g.hall@maths.abdn.ac.uk}, \mailto{DLonie@aol.com}}

\begin{abstract}
This paper discusses the extent to which one can determine the
space-time metric from a knowledge of a certain subset of the (unparametrised) geodesics
of its Levi-Civita connection, that is, from the experimental
evidence of the equivalence principle. It is shown that, if the
space-time concerned is known to be vacuum, then the Levi-Civita
connection is uniquely determined and its associated metric is
uniquely determined up to a choice of units of
measurement, by the specification of these geodesics. It is further
demonstrated that if two space-times share the same unparametrised
geodesics and {\it only one} is assumed vacuum then their
Levi-Civita connections are again equal (and so the other metric is
also a vacuum metric) and the first result above is recovered.
\end{abstract}
\pacs{04.20.-q, 04.20.Cv, 02.40.Ky }
\submitto{\CQG}
\maketitle
\section{Introduction}
    In Newtonian gravitational theory one can consider a particle as having, in principle, three types of mass; its active gravitational mass $m_{AG}$, its passive gravitational mass $m_{PG}$  and its inertial mass $m_{I}$.
    The mass $m_{AG}$ is a measure of the particle's ability to gravitationally attract another particle, whilst the mass $m_{PG}$ is a measure of its susceptibility to being gravitationally attracted by another particle.
    The mass $m_{I}$ is a measure of the particle's resistance to being accelerated in an inertial frame. Thus for two mutually attracting particles labelled $m$ and $M$ in an inertial frame and at distance $r$ apart, Newton's third and second laws give, respectively, for this two body problem
    \begin{equation}\label{1}
    M_{AG}m_{PG}=M_{PG}m_{AG}\qquad GM_{AG}m_{PG}=r^{2}m_{I}a
    \end{equation}
    where $G$ is the Newtonian gravitational constant and $a$ is the magnitude of the
    acceleration of m in this frame. From the first of these equations one gets
    $M_{AG}/M_{PG}=m_{AG}/m_{PG}$. It follows that one may choose units with which to measure the active and passive gravitational masses such that, for any particle, its active and passive gravitational masses are equal (and written as, say, $m_{G}$). The second equation in (\ref{1}) then shows that, within the gravitational field of $M$,  $aG^{-1}m_{I}/m_{G}$ is the same for all particles at a fixed event. Newtonian theory then assumes the constancy of $G$ and accepts the experimental result,
    contained within the principle of equivalence, that $a$ is the same for all particles at a fixed event. It follows that $m_{G}/m_{I}$  is particle independent and so, choosing
    appropriate units for the measurement of inertial mass, one may take for any particle $m_{G}=m_{I}$. The conclusion is that only one mass parameter is needed for each particle and that, from the linearity in Newtonian theory, a given gravitational field provides a well-defined gravitational acceleration at each event and which is inherited by each freely falling particle at that event, independently of its make-up. It follows that the path of a particle passing through that event depends only on its velocity at that event. This conclusion is one form of the principle of equivalence in Newtonian theory and appears as a consequence of Newton's laws together with the experimental results mentioned above. If one assumes from the outset the result that the inertial, active gravitational and passive gravitational masses are equal, then the constancy of the acceleration follows immediately from the constancy of $G$.

    In Einstein's general relativity theory, the (weak) principle of equivalence now arises, based on the experimental evidence above, as an assumption regarding the paths of such freely falling particles at some space-time event, this assumption amounting to their dependence, as in the Newtonian case, only on the particle velocity at that event. This path is related to the geometry of space-time by being assumed to be (part of) a timelike geodesic of the Levi-Civita connection associated with the space-time metric. [Regarding the extent to which this result can be proved from the other axioms of Einstein's theory, see \cite{1}.]

    This paper examines the extent to which one can identify the space-time metric in general relativity from a knowledge of a certain set of space-time paths representing such freely falling particles and which are assumed to be timelike geodesics. It generalises work in an earlier paper \cite{2} in which this problem was briefly considered and provides the details omitted in that paper. In this paper, to clarify notation, the term {\it geodesic} will be used in the most general sense with the curve parameter arbitrary. The term {\it unparametrised geodesic} is sometimes used in this sense. On the other hand, an {\it affinely parametrised geodesic} will be referred to as just that.

    Let $M$ be a smooth space-time manifold with all geometrical objects defined on $M$ smooth. Consider the following general situation. Let $g$ and $g{'}$ be Lorentz metrics on $M$ with signatures $(-,+,+,+)$ and with associated Levi-Civita connections $\nabla$ and $\nabla{'}$.
    Suppose that for each $p\in M$ there is an open subset $G_{p}\neq\emptyset$ of the tangent space $T_{p}M$ to $M$ at $p$ such that for each $p\in M$ and for each $u\in G_{p}$, $u$ is timelike with respect to $g$ and $g{'}$ and that there exists a curve in $M$ containing $p$ whose tangent at $p$ is $u$ and which is an unparametrised
    geodesic with respect to both $\nabla$ and $\nabla{'}$.
    How are $\nabla$ and $\nabla{'}$ related and how are $g$ and $g{'}$ related? In other words, given that the principle of equivalence determines local (unparametrised) geodesic paths for a certain family of particles at each $p\in M$, how much does it say about the Levi-Civita connection and metric on $M$? The assumptions made so far in this paragraph will be referred to as {\it assumption} $A$ in what is to follow. It is remarked here that $G_{p}$ may be interpreted for some given observer at $p$
    as the collection of all particle tangent vectors at $p$ for which the equivalence principle has been established.
    [In fact, in what is to follow, the condition that the members of
    $G_{p}$ are timelike is not used and so the initial parts of the argument
    in section 2 could be applied even if the connections in question were not
    metric.]

    From the physical viewpoint, one could argue that the internal motions of certain
    freely falling particles could identify a ${\it proper time}$ along
    such geodesic paths and hence a common affine parameter for
    both $\nabla$ and $\nabla{'}$. With this additional assumption, the geodesic
    equations give $\Gamma^{a}_{bc}u^{b}u^{c}$=$\Gamma{'}^{a}_{bc}u^{b}u^{c}$
    at $p$ for each $u\in G_{p}$, where $\Gamma$ and $\Gamma{'}$ represent the Christoffel symbols for $\nabla$ and $\nabla{'}$, respectively, in some (any) coordinate neighbourhood of $p$.
    It follows from this that, since $G_{p}$ is open in $T_{p}M$, $\Gamma^{a}_{bc}=\Gamma{'}^{a}_{bc}$ at $p$ and hence $\nabla$ and $\nabla{'}$ are identical connections on $M$.
    To see this, one notes that if $P^a_{bc}=\Gamma'^a_{bc}-\Gamma^a_{bc}$ then, in any coordinate system about $p$, $P^a_{bc}u^bu^c$ is a polynomial in the components of $u$ which vanishes on the open subset $G_p$ of $\R^4$. It follows that $P^a_{bc}(p)=0$ at $p$.
    From this equality of the connections $\nabla$ and $\nabla'$, the relationship between $g{'}$ and $g$ depends on the common {\it holonomy group} of $\nabla$ and $\nabla'$ and is easily found \cite{3,4}. In particular, it turns out that, {\it generically} \cite{4,5}, $g{'}=cg, (0<c\in \R)$ and so, generically, the metric is known up to a (positive) {\it constant} conformal factor, that is, up to the units of measurement. If the further assumption is made, that experiments with
    light rays determine the null cone at each point, then $g$ and $g{'}$ are necessarily ${\it always}$ conformally related and this, together with the above result $\nabla =\nabla{'}$, leads easily to the result that $g'=cg, (0<c\in \R)$.
    \section{Generalisations}
    {\it Suppose now that only assumption $A$ is made.} Then, using a semi-colon and a vertical stroke for the covariant derivatives with respect to $\nabla$ and
    $\nabla'$, respectively, one finds for the associated {\it unparametrised} geodesic equations for these connections and for a common geodesic $x^{a}(t)$ through any $p\in M$, $x^{a}_{\ ;b}x^{b}x^{e}=x^{e}_{\ ;b}x^{b}x^{a}$ and $x^{a}_{\ | b}x^{b}x^{e}=x^{e}_{\ | b}x^{b}x^{a}$ and hence the relations (cf. \cite{6,7}),
    \begin{equation}\label{2}
    (\delta^{b}_{\ c}P^{a}_{de}-\delta^{a}_{\ c}P^{b}_{de})u^{c}u^{d}u^{e}=0
    \end{equation}
    for each $u\in \bigcup_{p\in M}G_{p}$ and where, in (\ref{2}),  $P$ is as before. One can now show that the bracketed part of (\ref{2}), symmetrised over the indices $c,d$ and $e$, is zero.
    To see this, one simply extends the argument of the previous paragraph.
    Thus, on performing this symmetrisation, one finds the necessary and sufficient condition that $\nabla$ and $\nabla'$ have the same (unparametrised) geodesics (i.e. that they are {\it projectively related}) to be (c.f.\cite{6,7})
    \begin{equation}\label{3}
    P^a_{bc}=\Gamma'^{a}_{bc}-\Gamma^{a}_{bc}=\delta^{a}_{\ b}\psi_{c}+\delta^{a}_{\ c}\psi_{b}
    \end{equation}
    for some smooth global 1-form $\psi$ on $M$. Also, since $\nabla$ and $\nabla{'}$ are {\it metric} connections, $\psi$ is easily checked to be a ${\it closed}$ 1-form and hence is locally a gradient \cite{6}. From (\ref{3}) the condition that $\nabla{'}g'=0$ may be written in the equivalent form
    \begin{equation}\label{4}
    g{'}_{ab;c}=2g{'}_{ab}\psi_{c}+g{'}_{ac}\psi_{b}+g{'}_{bc}\psi_{a}
    \end{equation}
    In fact, (\ref{3}) and (\ref{4}) are equivalent conditions for projective relatedness. To see this write the difference between the $\nabla$ and $\nabla'$ covariant derivatives of $g'$ in terms of $P$ in an obvious way (and using
    (\ref{4})) and then permute the indices in the resulting (coordinate) equation in the way one normally does to obtain the expression for the Levi-Civita connection
    in terms of the metric. One thus obtains (\ref{3}). Equation (\ref{3}) reveals a simple relationship between the type (1, 3) curvature tensors $R$ and $R'$ arising from $\nabla$ and $\nabla{'}$, respectively, and which can be written in an obvious notation as \cite{6}
    \begin{equation}\label{5}
    R{'}^{a}{}_{bcd}=R^{a}{}_{bcd}+\delta^{a}_{\ d}\psi_{bc}-\delta^{a}_{\ c}\psi_{bd}
    \end{equation}
    where $\psi_{ab}=\psi_{a;b}-\psi_{a}\psi_{b} (=\psi_{ba}$). It is remarked that (\ref{5}) depends on the symmetry of $\psi_{ab}$, that is, on the metric condition on $\nabla$ and $\nabla{'}$ and the consequent fact that $\psi$ is closed. Otherwise, extra terms occur in (\ref{5}). From (\ref{5}), or more precisely from the fact that $\nabla$ and $\nabla'$ are projectively related, it can be checked that the {\it Weyl projective tensor} $W$, with components
    \begin{equation}\label{6}
    W^{a}{}_{bcd}=R^{a}{}_{bcd}-\case{1}{3}(\delta^{a}_{\ c}R_{bd}-\delta^{a}_{\ d}R_{bc})
    \end{equation}
    where $R_{ab} (=R^{c}{}_{acb})$ are the Ricci tensor components, is the same for $\nabla$ and $\nabla{'}$ and so the expression in (\ref{6}) is unchanged if the curvature and Ricci tensors are exchanged for their primed counterparts \cite{8}.

    So far the discussion has been quite general and based on assumption $A$. The following particular case was described briefly in \cite{2}.
    Suppose that $g$ and $g{'}$ are {\it each vacuum metrics}. Then the above remarks about the tensor $W$, together with (\ref{6}), show that, on $M$,
    \begin{equation}\label{7}
    R{'}^{a}{}_{bcd}=R^{a}{}_{bcd}
    \end{equation}
    Thus the type (1,3) curvature tensors of $g$ and $g'$ are equal on $M$. It then follows \cite{9,4} that the Petrov types of $g$ and $g{'}$ are the same at each $p\in M$. [It is remarked here that this is not completely obvious since the Petrov classification is an algebraic statement depending on the Weyl (or, in the vacuum case, the curvature) tensor {\it and} the metric and one needs to know that (\ref{7}) imposes a significant restriction on how $g$ and $g'$ are related.]
    Now the (common) Petrov type of $g$ and $g'$ may vary over $M$ and is important for determining this relationship between $g$ and $g'$.
    So consider the disjoint decomposition $M=H\cup \mb{N}\cup\mb{O}$ where $\mb{N}$ is
    the set of all points of $M$ at which the Petrov type is $\mb{N}$, $\mb{O}$ is the set of all points of $M$ at which the curvature tensor vanishes and $H$ is defined by the disjointness of the decomposition.
    Here, the physically reasonable {\it non-flat} assumption will be made that $\mathrm{int}\mb{O}$ is empty (where $\mathrm{int}$ denotes the interior operator in the manifold topology on $M$) so that the curvature tensor cannot vanish over any non-empty open subset of $M$. It is also noted, by a rank argument on the well-known $6\times 6$ symmetric matrix form of the (smooth and hence continuous) curvature tensor, that $H$ is an open subset of $M$ since the points of $H$ are exactly the points of $M$ where this matrix rank is at least four (see, for example, \cite{4}) the rank being the same whether the curvature tensor is taken in its (tensor) type $(0,4)$ or $(2,2)$ form.
    Then write the {\it disjoint} decomposition $M=H\cup \mathrm{int}\mb{N}\cup \mb{O}\cup P$ where $P=\mb{N}\setminus \mathrm{int}\mb{N}$. Now $\mb{O}\cup P$ is a closed subset of $M$, being the complement in $M$ of the open subset $H\cup \mathrm{int}\mb{N}$.
    Also $H\cup \mb{N}$ is open in $M$ and so $\mb{O}\cup P$ has empty interior in
    $M$. This follows since if $U$ is an open subset of $\mb{O}\cup P$ then the non-empty open subset $V\equiv H\cup \mb{N}$ satisfies the condition that $U\cap V$ is an open subset of $P$ and hence empty, by the definition of $P$. So, by definition of $V$,
    $U\subset\mb{O}$ and so $U$ is empty by the non-flat condition. Thus $\mb{O}\cup P$ is closed with empty interior in $P$.
    This shows that $\mb{O}\cup P$ is {\it nowhere dense} in $M$ and hence that $H\cup \mathrm{int}\mb{N}$ is an {\it open dense} subset of $M$.

    Now suppose $H$ is not empty. Then, because of (\ref{7}) and the above mentioned curvature rank condition on $H$, $g$ and $g'$ are conformally related with a constant conformal factor on any open connected subset $U$ of $H$ \cite{9,4}, and so $g=\rho g'$, with  $\rho$ constant on $U$. Thus $\nabla=\nabla{'}$ on $U$ and hence on $H$.
    If $\mathrm{int}\mb{N}$ is not empty, then for each $p\in \mathrm{int}\mb{N}$, there is a connected open neighbourhood $V$ of $p$ and a nowhere zero {\it smooth} (see \cite{16}) null vector field $l$ on $V$ spanning the repeated principal null direction of the curvature(s) at each point of $V$ (that is, $R^{a}{}_{bcd}l^{d}=0$ on $V$) such that, on $V$, \cite{9,4}
    \begin{equation}\label{8}
    g{'}_{ab}=\phi g_{ab}+\alpha l_{a}l_{b}\qquad (l_{a}\equiv g_{ab}l^{b})
    \end{equation}
    for functions $\phi$ and $\alpha$ :$V \rightarrow\R$ with $\phi$ positive and which are easily checked to be smooth since $g$, $g'$ and $l$ are. Further, (\ref{5}) and (\ref{7}) together with a contraction over the indices $a$ and $c$ reveal that $\psi _{ab}$=0 and so $\psi_{a;b}=\psi_{a}\psi_{b}$ on $V$.
    Now suppose $\psi(p)\neq 0$ and hence, by reducing $V$ if necessary, that $\psi$ is nowhere zero on $V$.  Now since $\psi_{a;b}=\psi_{a}\psi_{b}$, the 1-form $\psi$ is ${\it recurrent}$ (with respect to $\nabla$) and nowhere zero on $V$. This recurrence property, together with the Ricci identity for $\psi$, can be used to show that $\psi^{a}\equiv g^{ab}\psi_{b}$ is a principal null direction of the curvature tensor (i.e. $R^a{}_{bcd}\psi^{d}=0$) and hence, by the uniqueness property of such directions, $\psi_{a}=\beta l_{a}$ for some smooth real valued nowhere zero function $\beta$ on $V$.
    Thus $l_{a}$ is {\it recurrent on} $V$ with respect to $\nabla$ (that is, $l_{a;b}=l_{a}q_{b}$ for some smooth 1-form $q$ on $V$). The Ricci identity for $l$ then shows that $q_{[a;b]}=0$ on $V$ and so, again by reducing $V$ if necessary, $q_{a}=\sigma,_{a}$ for some smooth function $\sigma:V\rightarrow \R$, where a comma denotes a partial derivative.
    Then $l'\equiv e^{-\sigma}l$ is a covariantly constant null vector field on $V$. Now define a function $\alpha'\equiv e^{2\sigma} \alpha$ on $V$ and rewrite (\ref{8}) as $g'_{ab}=\phi g_{ab}+\alpha'l'_{a}l'_{b}$. On substituting this last equation into (\ref{4}) one finds that $\phi_{,a}=\case{5}{2}\phi\psi_{a}$. A back substitution and an elementary rank argument then reveals that $\psi _{a}=\phi _{,a}=0$.
    This contradicts the fact that $\psi$ is non-zero on $V$ and so $\psi$ {\it must vanish identically on} $V$ and hence on $\mathrm{int}\mb{N}$.
    It follows from (\ref{3}) that $\nabla=\nabla'$ on $\mathrm{int}\mb{N}$ and hence (from the previous argument) on the open dense subset $H\cup \mathrm{int}\mb{N}$ of $M$. Thus the Levi-Civita connections $\nabla$ and $\nabla'$ on $M$ associated with the metrics $g$ and $g'$ on $M$ are identical.

    It has thus been established that $g$ and $g'$ have the same Levi-Civita connection and hence the same holonomy group. Since each is a vacuum metric this holonomy group is severely restricted and it follows \cite{10,4}, in the notation of these references, that the holonomy algebra is either of type $R_{8}$, $R_{14}$ or $R_{15}$. Further, if $H$ is not empty, it follows from the rank condition at points of $H$ and the fact that the infinitesimal holonomy algebra is contained in the holonomy algebra, that this algebra cannot be of type $R_{8}$ and so must be of
    type $R_{14}$ or $R_{15}$. Thus \cite{3,4}, if $H$ is not empty, $g$ and $g'$ {\it are conformally related on $M$ with a constant conformal factor}. If $H$ is empty, $M=\mb{N}\cup O$ with $\mb{N}$ open and dense in $M$. In this case, if the common holonomy type of $\nabla$ and $\nabla'$ is $R_{14}$ or $R_{15}$, one again has $g'=cg$ on $M$ for some constant $c\in \R$. If this holonomy type is $R_{8}$ then, if $M$ is {\it simply connected}, $M$ admits a global covariantly constant null vector field $l$ (with respect to $g$ and $g'$) whose direction coincides with the (unique) repeated principal null direction of the curvature at points of $\mb{N}$. Then $g$ and $g'$ are related by (\ref{8}) on $M$ with $l$ as in the previous sentence and $\phi$ and $\alpha$ constants and $M$, with either metric $g$ or $g'$, being a {\it pp-wave} space-time.
    If it is not simply connected, perhaps the best possible is to settle for a local representation of the relation between $g$ and $g'$ in some simply connected neighbourhood $V$ of any point of $\mathrm{int}\mb{N}$ of the form (\ref{8}) with $l$ covariantly constant on $V$ and $\phi$ and $\alpha$ constants.
    Since $M$ is connected, the smooth function $\chi\equiv g^{ab}g'_{ab}$ on $M$ is constant and coincides with $4\phi$ for each of these local representations. The constant $\alpha$ depends on the local representative vector field $l$ chosen for the principle direction.

    In conclusion, these results show that under the assumption $A$, together with the assumption that {\it both} $g$ and $g'$ are non-flat vacuum metrics, $g'$=$cg \,(0<c\in\R)$ on $M$ except in the special cases described above where the Petrov type is $\mb{N}$ or $\mb{O}$ everywhere. It is remarked that, always, $\nabla=\nabla{'}$ on $M$ and so if one of the geodesics considered is affinely parametrised with respect to $\nabla$, it is automatically affinely parametrised with respect to $\nabla{'}$.
    It is noted that the null cones of $g$ and $g'$ agree on $M$, except in these special cases, where the repeated principal null direction is their only common null direction. In geometrical terms (and considering the (precisely defined; see \cite{5}) generic case where these special cases have been removed) it says that each equivalence class of projectively related connections on $M$ contains at most one which is
    a Levi-Civita connection of a vacuum metric and this metric is then determined up to a constant conformal factor.

    Now suppose that, in addition to assumption A, the physical assumption regarding ${\it null}$ geodesics is introduced (see the last sentence of section 1) so that the null cones of $g$ and $g{'}$ coincide. Then $g{'}=\chi g$ on $M$, where $\chi:M\rightarrow \R$ is a positive function, and {\it with no assumption whatsoever on the energy-momentum tensor}, one can show, by a substitution into (\ref{4}) and a simple rank argument similar to one given above, that $\chi_{,a}=2\chi\psi_a$. A back substitution then shows that $\psi=0$ on $M$ and hence that $\chi$ is constant on $M$.
   Thus $\nabla=\nabla{'}$ on $M$ and an affine parameter with respect to $\nabla$
   is also affine with respect to $\nabla{'}$, and ${\it vice versa}$. [The authors have since found out that this last result is known \cite{7}]. This result reflects the fact that a vector field on M which is simultaneously projective and conformal is homothetic \cite{4}.
   \section{The General Case}
   In the next two sections, space-times with the same geodesics will again be considered but this time {\it only one of the space times will be assumed to be a vacuum space-time}. It will be shown that the other one {\it must} be vacuum and that the two metrics are then related as in the previous section. This result means that if a bunch of (unparametrised) geodesics of a space-time manifold are given (in the sense made precise in section 1) there is at most one Levi-Civita connection that is simultaneously compatible with a vacuum metric and consistent with these geodesics and, apart from the type $\mb{N}$ possibilities given earlier, the metric is uniquely
   determined up to a constant conformal factor (that is, up to units of measurement).

   So suppose, as before, that M is a space-time manifold and that $g$ and $g'$ are Lorentz metrics on $M$ with respective Levi-Civita connections $\nabla$ and $\nabla'$. Let assumption $A$ of section 1 hold and suppose, in addition, that $g$ is a {\it vacuum} metric. How are $g$ and $g'$ related? It follows from the work in section
   2 that $\nabla$ and $\nabla'$ are projectively related, that (\ref{3}) and (\ref{4}) hold and that the associated curvature tensors $R$ and $R'$ satisfy (\ref{5}) with $R^{a}{}_{bcd}$ the components of a {\it vacuum} curvature tensor (so that $R_{ab}=0$). The properties of the Weyl projective tensor $W$ given in (\ref{6})
   then reveal that (\ref{7}) is replaced by
   \begin{equation}\label{9}
   R^{a}{}_{bcd}=R{'}^{a}{}_{bcd}-1/3(\delta^{a}_{c}R'_{bd}-\delta^{a}_{d}R'_{bc})
   \end{equation}
   where $R'_{ab}=R'^{c}{}_{acb}$. One now attempts to deduce the restrictions that regulate the relationship between $g$ and $g'$ from (\ref{4}) and the consequent relation (\ref{9}) on the curvature tensors.

   To do this it is convenient to work with the vacuum curvature tensor $R$ which, being equal to the Weyl {\it conformal} tensor (not to be confused with the Weyl projective tensor $W$), satisfies the self dual condition $\overset{*\ \ }{R}=\overset{\ \ \ *}{R}(\Leftrightarrow\ \overset{*\ \ *}{R}=-R)$ where a star denotes the Hodge duality operator. This means that at each point of $M$, $R$ may be written as one of the canonical (algebraic) Petrov types \cite{11,12,13,14}. It is also noted that $M$ admits {\it disjoint} decompositions of the form \cite{4}
   \begin{equation}\label{10}
   M=\mb{I}\cup \mb{II}\cup \mb{D}\cup \mb{III}\cup
   \mb{N}\cup \mb{O}
   \end{equation}
   \begin{equation}\label{11}
   M=\mb{I}\cup \mathrm{int}\mb{II}\cup \mathrm{int}\mb{D}\cup
   \mathrm{int}\mb{III}\cup \mathrm{int}\mb{N}\cup E
   \end{equation}
   where $\mb{I}$ denotes the subset of $M$ of all points where the Petrov type is $\mb{I}$ and similarly for the other Petrov types $\mb{II}$, $\mb{D}$, $\mb{III}$, $\mb{N}$ and $\mb{O}$ (no confusion will arise from this ambiguity of notation), where $E$ is a closed subset of $M$ defined by the disjointness of the decomposition, where $\mb{I}$ is necessarily open and where, retaining the non-flat condition that
   $\mathrm{int}\mb{O}$ is empty, $\mathrm{int}E$ is empty. Thus $E$ is nowhere dense in $M$ and $M\setminus E$ is open and dense in $M$. [The breakdown of M into the interiors of the subsets
   $\mb{I}$, $\mb{II}$, $\mb{D}$, $\mb{III}$ and $\mb{N}$ is to allow calculus
   to be performed on them.]
   The procedure is then to use the individual Petrov types for $R$ to get an algebraic relation between $g$ and $g'$ at each point of each Petrov type neighbourhood of $M$ and then to finally impose the projective condition. Three lemmas are first required which are quite general (that is, independent of the Petrov type). Their relevance stems from the following result concerning the algebraic structure of the vacuum
   curvature tensor $R$. The Petrov type can be described entirely in terms of the {\it complex self dual eigenbivectors} of the {\it complex self dual vacuum curvature tensor} $\overset{+\ \ }{R}=R+\rmi \overset{*\ \ }{R}$ (curvature eigenbivectors).
   Thus if $\overset{+}{F}$ is a complex self dual curvature eigenbivector, so that $\overset{+}{F}=F+\rmi \overset{*}{F}$ for some real bivector $F$, then $\overset{+\ \ }{R}{}^{ab}{}_{cd}\overset{+}{F}{}^{cd}=2z\overset{+}{F}{}^{ab}$ where the {\it eigenvalue}
   $z=\alpha+\rmi \beta\in \C$ ($\alpha,\beta\in\R$), the factor $2$ being introduced for later convenience.
   All indices are raised and lowered using the metric $g$ unless specified otherwise.
   More conveniently, use of the self dual condition on the curvature tensor shows this last statement to be equivalent to the statement that
   $R^{ab}{}_{cd}\,\,\overset{+}{F}{}^{cd}=z\overset{+}{F}{}^{ab}$ (and hence the convenience of the factor 2
   mentioned above). Thus one can work equivalently, and more conveniently, with the real curvature tensor.
   Now if, in addition,  $\overset{+}{F}$ is {\it null}, $F$ and $\overset{*}{F}$ are each simple. If $\overset{+}{F}$ is {\it non-null}, a duality rotation of $\overset{+}{F}$, given by
   $\overset{+}{F}\rightarrow e^{\rmi \theta}\overset{+}{F}$ with $\theta\in \R$, may be chosen so that the real part of the (new) curvature eigenbivector $e^{\rmi \theta}\overset{+}{F}$ is simple and timelike (respectively spacelike) and its imaginary part (dual) is simple and spacelike (respectively timelike) (see e.g. \cite{4}). Thus for any complex self dual curvature eigenbivector $\overset{+}{F}$, one may arrange, by a suitable (complex) scaling,
   \begin{equation}\label{12}
   R^{ab}_{\ \ cd}F^{cd}=\alpha F^{ab}-\beta\overset{*}{F}{}^{ab} \qquad
   R^{ab}_{\ \ cd}\overset{*}{F}{}^{cd}=\alpha\overset{*}{F}{}^{ab}+\beta F^{ab}
   \end{equation}
   with $F$ (and hence $\overset{*}{F}$) simple. Another relationship which is useful follows from the identity
   $g'_{ea}R'^{e}{}_{bcd}+g'_{eb}R'^{e}{}_{acd}=0$ together with (\ref{9}) and is
   \begin{equation}\label{13}
   g'_{ae}R^{e}{}_{bcd}+g'_{be}R^{e}{}_{acd}=-\case{1}{3}(g'_{ac}R'_{bd}+g'_{bc}R'_{ad}-g'_{ad}R'_{bc}-g'_{bd}R'_{ac})
   \end{equation}
   To prepare for the lemmas mentioned above, $l, n, x, y$ will denote a real null tetrad (so that the only non-vanishing inner products between its members are
   $l^{a}n_{a}=x^{a}x_{a}=y^{a}y_{a}=1$) and $u, x, y, z$ will denote an associated pseudo-orthonormal tetrad  with $u=2^{-1/2}(l-n)$ and $z=2^{-1/2}(l+n)$ (so that the only non-vanishing inner products between its members are
   $-u^{a}u_{a}=x^{a}x_{a}=y^{a}y_{a}=z^{a}z_{a}=1$). For each tetrad member, it is convenient to have a covector associated with it through $g'$ in addition to the natural one obtained from $g$. Thus the covector $u'$ is defined by
   $u'_{a}\equiv g'_{ab}u^{b}$ and similarly for the other tetrad members. Also, if $F$ is a simple bivector, say $F=r^{a}s^{b}-s^{a}r^{b}$ (written $r\wedge s$) with $r$ and $s$ independent members of the tangent space $T_{p}M$, the 2-dimensional subspace (2-space) of $T_{p}M$ spanned by $r$ and  $s$ (and which is uniquely determined by $F$) is called the {\it blade} of $F$. It is finally remarked that the individual Petrov types can be characterised by their eigenbivector structure and this will be introduced when required. In particular, the {\it algebraically special} types can be characterised by the existence of a
   {\it null} complex self dual eigenbivector at the appropriate point $p\in M$. This can be seen either by inspecting the usual Petrov canonical forms or by the following argument. Suppose (\ref{12}) holds at $p$   for a null bivector $F$ with principal null direction $l$ (so that for some null tetrad $l, n, x, y$ at $p$,
   $F_{ab}=l\wedge x$ and $\overset{*}{F}_{ab}=-l\wedge y$). Then define the symmetric tensor $T$ at $p$ by $T_{bd}=R_{abcd}l^{a}l^{c}$, from which it follows that
   $T_{ab}l^{b}=T_{ab}x^{b}=T_{ab}y^{b}=0$. Thus $T_{ab}\propto l_{a}l_{b}$ which is the (Bel \cite{14}) condition that the curvature tensor is algebraically special with
   (repeated) principal null direction $l$ at $p$. Conversely, if this last (algebraically special Bel) condition holds on the curvature tensor at $p$, then, on extending $l$ to a null tetrad $l,n,x,y,$ at $p$, it is easily checked that the dual pair of bivectors $G_{ab}\equiv R_{abcd}l^{c}x^{d}$ and $\overset{*}{G}_{ab}=-R_{abcd}l^{c}y^{d}$ satisfy $G_{ab}l^{b}=\overset{*}{G}_{ab}l^{b}=0$. It follows that they are null bivectors with common principal null direction $l$ and hence are independent linear combinations of the bivectors $F\equiv l\wedge x$ and $\overset{*}{F}\equiv -l\wedge y$. Thus $F$ and $\overset{*}{F}$ satisfy (\ref{12}) at $p$ and the existence of a
   complex self dual curvature eigenbivector at $p$ is assured.

   Thus for each of the following lemmas, $l,x,y,z$ and $u,x,y,z$ are the tetrads given (and related) above, and $g$ and $g'$ are Lorentz metrics on $M$ with Levi-Civita connections $\nabla$ and $\nabla'$ and curvature tensors $R$ and $R'$, respectively.
   The Levi-Civita connections are supposed controlled by assumption A as described in section 1 and hence are projectively related.
   In addition, $g$ is a non-flat {\it vacuum} metric and $\overset{+}{F}$ is a non-zero complex self dual curvature eigenbivector of the vacuum curvature tensor $R$ at $p\in M$ with eigenvalues $\alpha+\rmi \beta$. The notation for the covectors $u'$, etc, is as given earlier. It is added that statements like $k$ is an eigenvector of, for example, $g'$ with eigenvalue $\lambda$ are understood to mean with respect to $g$, that is, $g'_{ab}k^b=\lambda g_{ab}k^b(=\lambda k_a)$.
   \begin{Lemma}
   At $p$ let $\overset{+}{F}$ be {\it non-null} with associated
   eigenvalue $\alpha+\rmi \beta$ so
   that (\ref{12}) holds with
   $F=u\wedge z=l\wedge n$ and $\overset{*}{F}=x\wedge
   y$ for a suitable tetrad (so that $F$ is timelike and $\overset{*}{F}$ spacelike).
\begin{enumerate}[(i)]
\item If $\beta\neq0$, then, at $p$ with indices omitted, so that
$u$ means $u_a$ etc. and, to avoid confusion with the curvature tensor $R'$, terms like $R'_{ab}u^b$ will be denoted by $\tilde{R}u$, etc.
\begin{equation}\label{14}
\fl\begin{array}{lllll}
\beta u&=&b_{00}u'\hspace{13mm}-b_{02}x'-b_{03}y' \\
\beta z&=&\hspace{13mm}b_{00}z'+b_{12}x'+b_{13}y' \\
\beta x&=&b_{02}u'+b_{12}z'+b_{22}x'\hspace{13mm}&=&c_3l'+c_1n'+b_{22}x'\hspace{13mm}\\
\beta y&=&b_{03}u'+b_{13}z'\hspace{13mm}+b_{22}y'&=&c_4l'+c_2n'\hspace{13mm}+b_{22}y'\\
\beta l&=& & &b_{00}l'\hspace{13mm}+c_1x'+c_2y' \\
\beta n&=& & &\hspace{11mm}b_{00}n'+c_3x'+c_4y' \\
\end{array}
\end{equation}
\begin{equation}\label{15}
\fl\begin{array}{lllll}
\alpha u+\case{2}{3}\tilde{R}u&=&a_{00}u'\hspace{13mm}-b_{13}x'+b_{12}y' \\
\alpha z+\case{2}{3}\tilde{R}z&=&\hspace{13mm}a_{00}z'+b_{03}x'-b_{02}y' \\
\alpha x+\case{2}{3}\tilde{R}x&=&b_{13}u'+b_{03}z'+a_{22}x'\hspace{13mm}&=&c_4l'-c_2n'+a_{22}x'\hspace{13mm}\\
\alpha y+\case{2}{3}\tilde{R}y&=&-b_{12}u'-b_{02}z'\hspace{10mm}+a_{22}y'&=&-c_3l'+c_1n'\hspace{10mm}+a_{22}y'\\
\alpha l+\case{2}{3}\tilde{R}l&=& & &a_{00}l'\hspace{13mm}-c_2x'+c_1y' \\
\alpha n+\case{2}{3}\tilde{R}n&=& & &\hspace{11mm}a_{00}n'+c_4x'-c_3y' \\
\end{array}
\end{equation}
   where the coefficients $a_{ij}$, $b_{ij}$ and $c_i$ are real numbers.
\item If $\beta=0$, then, at $p$
\begin{equation}\label{16}
\begin{array}{lll}
\alpha u+\case{2}{3}\tilde{R}u&=&au' \\
\alpha z+\case{2}{3}\tilde{R}z&=&az' \\
\alpha x+\case{2}{3}\tilde{R}x&=&bx' \\
\alpha y+\case{2}{3}\tilde{R}y&=&by' \\
\alpha l+\case{2}{3}\tilde{R}l&=&al' \\
\alpha n+\case{2}{3}\tilde{R}n&=&an' \\
\end{array}
\end{equation}
with $a,\,b\in \R$ and so the 2-dimensional subspaces of $T_{p}M$ spanned by $u$ and $z$ and also by $x$ and $y$ are eigenspaces of the tensors $ag'_{ab}-\case{2}{3}R'_{ab}$ and $bg'_{ab}-\case{2}{3}R'_{ab}$, respectively, with eigenvalue $\alpha$.
\end{enumerate}
\end{Lemma}
\begin{Lemma}
At $p$ let $\overset{+}{F}$ be {\it null} with associated eigenvalue $\alpha+\rmi \beta$ so that (\ref{12}) holds with $F=l\wedge x$ and $\overset{*}{F}=-l\wedge y$ for a suitable tetrad.
\begin{enumerate}[(i)]
\item If $\beta\neq 0$, then, at $p$, $l$ is an eigenvector of $R'_{ab}$ and also of $g'_{ab}$

\item If $\beta=0$, there exists $\delta\in\R$ such that the 3-dimensional
subspace of $T_pM$ spanned by $l$, $x$ and $y$ is an eigenspace of
$\delta g'_{ab}-\case{2}{3}R'_{ab}$ with eigenvalue $\alpha$.
 \end{enumerate}
 \end{Lemma}
    \begin{Lemma}
    Let $\overset{+}{F}$ and $\overset{+}{G}$ be any (null or non-null) complex self dual curvature eigenbivectors at $p$ with identical real eigenvalues (so that part (ii) of lemmas $1$ and $2$ hold). If any subset of the blades of the real and imaginary parts of $\overset{+}{F}$ and $\overset{+}{G}$ (chosen simple as described earlier) has a non-zero tangent vector at $p$ in common, the span of this set of blades is an eigenspace of $\delta g'_{ab}-2/3R'_{ab}$ at $p$ for some $\delta \in \R$ with eigenvalue $\alpha$.
   \end{Lemma}

   What lemmas $1(ii)$ and $2(ii)$ say is that if a curvature eigenbivector $\overset{+}{F}$ at $p$ , either null or non-null, has an associated real eigenvalue $\alpha$ then $F$ and $\overset{*}{F}$ (which may be chosen simple and are also curvature eigenbivectors with real eigenvalue $\alpha$) are such that their blades are eigenspaces of $R_{ab}-\delta g_{ab}$ at $p$ for some $\delta\in\R$ which is dependent on the blade. In the null case, the null vector $l$ lies in the intersection of the blades and the corresponding numbers $\delta$ are equal. Lemma $3$ takes care of a generalised version of this
   result.

   $\mb{Proof}$
   For lemma 1 one contracts (\ref{13}) with $F^{cd}=u^{c}z^{d}-z^{c}u^{d}$, replacing the terms which arise on the left hand side and which contain the curvature tensor $R$ by the first expression in (\ref{12}). One obtains
   \begin{equation}\label{17}
   \fl\hspace{1cm}\begin{array}{r}3\alpha (g'_{ae}F^e_{\ b}+g'_{be}F^e_{\ a})-3\beta
   (g'_{ae}\overset{*}{F}{}^e_{\ b}+g'_{be}\overset{*}{F}{}^e_{\ a})= \hspace{2cm}\\
   g'_{ae}F^{ce}R'_{bc}+g'_{be}F^{ce}R'_{ac}-g'_{ae}F^{ed}R'_{bd}-g'_{be}F^{ed}R'_{ad}
   \end{array}\end{equation}
   Equation (\ref{17}) can then be rewritten using the definitions $u'_{a}\equiv g'_{ab}u^{b}$, etc, given earlier, together with the abbreviations $\mu_{a}\equiv
   R'_{ab}u^{b}$ and $\nu_{a}\equiv R'_{ab}z^{b}$ as
   \begin{equation}\label{18}
   \fl\hspace{1cm}\begin{array}{r}3\alpha(u'_{a}z_{b}-z'_{a}u_{b}+u'_{b}z_{a}-z'_{b}u_{a})-3\beta(x'_{a}y_{b}-y'_{a}z_{b}+x'_{b}y_{a}-y'_{b}x_{a})\\
   =2(z'_{a}\mu_{b}+\mu_{a}z'_{b}
   -u'_{a}\nu_{b}-\nu_{a}u'_{b})
   \end{array}
   \end{equation}
   This simplifies to
   \begin{equation}\label{19}
   u'_{a}p_{b}+p_{a}u'_{b}-z'_{a}q_{b}-q_{a}z'_{b}
   =\beta(x'_{a}y_{b}+y_{a}x'_{b}-x_{a}y'_{b}-y'_{a}x_{b})
   \end{equation}
   where $p_{a}\equiv \alpha z_{a}+2/3\nu_{a}$ and $q_{a}\equiv \alpha u_{a}+2/3\mu_{a}$. Similarly, one can repeat this calculation using the second equation in (\ref{12}) . Since no use was made of the (timelike or spacelike) nature of the members
   of the basis $u,x,y,z$, one merely swaps $u,x,y,z$ for $x,y,u,z$, respectively, and replaces $\beta$ by $-\beta$ in the above calculation. The equation corresponding to (\ref{19}) is
   \begin{equation}\label{20}
   x'_{a}r_{b}+r_{a}x'_{b}-y'_{a}s_{b}-s_{a}y'_{b}=-\beta(u'_{a}z_{b}+z_{a}u'_{b}-u_{a}z'_{b}-z'_{a}u_{b})
   \end{equation}
   where $r_{a}\equiv \alpha y_{a}+2/3R'_{ab}y^{b}$ and $s_{a}\equiv \alpha
   x_{a}+2/3R'_{ab}x^{b}$.

   In order to obtain the required information from the above
   equations certain expansions are required and which benefit from
   a temporary notational change in order to allow a summation notation to be
   used. First, denote the members of the basis $u, z, x, y$ of $T_{p}M$,
   noting the order change, by $e_{0}, e_{1}, e_{2}$ and $e_{3}$,
   respectively. Then denote the corresponding covector basis at $p$
   obtained from them and the metric $g$ at p (lowering indices) by
   $\tilde{e_{0}}$, $\tilde{e_{1}}$, $\tilde{e_{2}}$ and
   $\tilde{e_{3}}$ and the covector basis at $p$ similarly obtained from
   the metric $g'$ at $p$ by $e'_{0}$, $e'_{1}$, $e'_{2}$ and
   $e'_{3}$ ($=u', z', x'$ and  $y'$) . Then, at $p$, $g'$ may be expanded as the tensor
   product $g'=\sum c_{ab}\tilde{e_{a}}\otimes \tilde{e_{b}}$ where
   $C\equiv (c_{ab})$ are the components of a real symmetric non-singular
   matrix. Successive contractions of this expansion by the $e_{a}$
   reveal that $e'_{a}=\sum \eta _{ab}c_{bc}\tilde{e_{c}}$ and hence,
   on inverting, that $\tilde{e_{a}}=\sum d_{ab}\eta _{bc}e'_{c}$,
   where $D\equiv (d_{ab})$ is $C^{-1}$. Thus, $\tilde{e_{a}}=\sum
   b'_{ab}e'_{b}$ where $b'_{\alpha \beta}=b'_{\beta \alpha}$ and
   $b'_{0\alpha}=-b'_{\alpha 0}$ $(1\leq \alpha, \beta \leq3)$.
   These relations between the entries of the matrix components
   $b_{a b}$ will be useful in what is to follow. [In fact, one may
   write a ``completeness relation'' for $g'$ as
   $g'=-\tilde{e_{0}}e'_{0}+\tilde{e_{1}}e'_{1}+\tilde{e_{2}}e'_{2}
   +\tilde{e_{3}}e'_{3}=-e'_{0}\tilde{e_{0}}+e'_{1}\tilde{e_{1}}+e'_{2}\tilde{e_{2}}
   +e'_{3}\tilde{e_{3}}$.] Similarly, one can expand the symmetric
   tensor $S_{ab}\equiv \alpha g_{ab}+2/3R'_{ab}$ at $p$ and get
   expressions in terms of the basis $e'_{a}$ at $p$ for the
   covectors $p_{a}(=S_{ab}z^{b})$, $q_{a}(=S_{ab}u^{b})$,
   $r_{a}(=S_{ab}y^{b})$ and $s_{a}(=S_{ab}x^{b})$. However, there
   may not be relations on the entries in the expansions for $p, q,
   r$ and $s$ corresponding to those on the entries $b'_{ab}$. A
   final remark is that the above expansion for the $e_{a}$ is
   only used in the case $\beta \neq 0$ and, in fact, it is then
   convenient to expand the basis $\beta \tilde{e_{a}}\equiv \sum
   b_{ab}e'_{b}$ with $b_{ab}\equiv \beta b'_{ab}$.

   Thus in the case $\beta\neq 0$ at $p$, the expansions are $\beta\tilde{e_{a}}=\sum b_{ab}e'_{b}$ and, for $q, p, s$ and $r$ (in that order)
\begin{equation}\label{21}
   \begin{array}{lll}
   q=\alpha
   \tilde{e_{0}}+\case{2}{3}\tilde{R}u&=&a_{00}u'+a_{01}z'+a_{02}x'+a_{03}y'\\
   p=\alpha
   \tilde{e_{1}}+\case{2}{3}\tilde{R}z&=&a_{10}u'+a_{11}z'+a_{12}x'+a_{13}y'\\
   s=\alpha
   \tilde{e_{2}}+\case{2}{3}\tilde{R}x&=&a_{20}u'+a_{21}z'+a_{22}x'+a_{23}y'\\
   r=\alpha
   \tilde{e_{3}}+\case{2}{3}\tilde{R}y&=&a_{30}u'+a_{31}z'+a_{32}x'+a_{33}y'
   \end{array}
\end{equation}
    where, as mentioned earlier, $\tilde{R}u$ denotes the covector at $p$ with components $R'_{ab}u^{b}$ and similarly for $\tilde{R}z$, $\tilde{R}x$ and $\tilde{R}y$. Recalling that in this notation, $\tilde{e_{a}}$ are,    respectively, the covectors with components $u_{a}, z_{a}, x_{a}$ and $y_{a}$, one can substitute these expansions into (\ref{19}) and (\ref{20}) and equate coefficients of the $e'_ae'_b$ in an obvious way. One finds that several of these vanish and certain others are forced to be equal. Reverting to the original notation (and where the symbols $u,x,y,z$ will also be used to denote the covectors $\tilde{e}_a$ at $p$), one finally has, at $p$,
   \begin{equation}\label{22}
\begin{array}{lll} \beta u&=&b_{00}u'\hspace{13mm}-b_{02}x'-b_{03}y' \\
\beta z&=&\hspace{13mm}b_{00}z'+b_{12}x'+b_{13}y' \\
\beta x&=&b_{02}u'+b_{12}z'+b_{22}x'\hspace{13mm}\\
\beta y&=&b_{03}u'+b_{13}z'\hspace{13mm}+b_{22}y'
\end{array}
\end{equation}
\begin{equation}\label{23}
\begin{array}{lll}
q=\alpha u+\case{2}{3}\tilde{R}u&=&a_{00}u'\hspace{13mm}-b_{13}x'+b_{12}y' \\
p=\alpha z+\case{2}{3}\tilde{R}z&=&\hspace{13mm}a_{00}z'+b_{03}x'-b_{02}y' \\
s=\alpha x+\case{2}{3}\tilde{R}x&=&b_{13}u'+b_{03}z'+a_{22}x'\hspace{13mm}\\
r=\alpha y+\case{2}{3}\tilde{R}y&=&-b_{12}u'-b_{02}z'\hspace{13mm}+a_{22}y'
\end{array}
\end{equation}
This is the result claimed in lemma $1(i)$
   in terms of the bases $u,z,x,y$ and $u',z',x',y'$. Then, noting that $u=\case{1}{\sqrt{2}}(l-n)$ and $z=\case{1}{\sqrt{2}}(l+n)$, the remainder of the result in lemma $1(i)$ (i.e. in terms
   of the bases $l,n,x,y$ and $l',n',x',y'$) is obtained using straightforward linear combinations of the equations in (22) and (23). In the statement of lemma 1(i), $c_1=\case{1}{\sqrt{2}}(b_{12}-b_{02})$, $c_2=\case{1}{\sqrt{2}}(b_{13}-b_{03})$, $c_3=\case{1}{\sqrt{2}}(b_{12}+b_{02})$ and $c_4=\case{1}{\sqrt{2}}(b_{13}+b_{03})$.
If $\beta=0$ then the
   right hand side of (\ref{19}) is zero. Thus $z'$ and $q$ are linear
   combinations of $u'$ and $p$ in the cotangent space $T_{p}^{\ *}M$ to $M$ at $p$.
   So by writing $z'=\gamma u'+\delta p$ and $q=\gamma 'u'+\delta
   'p$ ($\gamma, \delta, \gamma',\delta'\in \R$) and substituting
   into (\ref{19}) one finds that $\gamma=\delta'=0$ and that $\gamma'
   \delta=1$. Thus $\delta\neq 0$, $z'=\delta p$ and
   $q=\delta^{-1}u$. These are the first two claimed results of
   lemma $1(ii)$ with $\delta^{-1}=a$.
   Again,
   noting that $u=\case{1}{\sqrt{2}}(l-n)$ and $z=\case{1}{\sqrt{2}}(l+n)$, the last two equations in (16) of lemma $1(ii)$ follow immediately.
   A similar analysis on (\ref{20}) reveals the other two results and completes the proof of lemma $1(ii)$ and hence of lemma $1$.

   For lemma $2$ one starts by writing $F=l\wedge x$ and
   $\overset{*}{F}=-l\wedge y$. Then (\ref{13}) is contracted with $F^{cd}$ and terms
   involving the curvature replaced by those on the right hand side
   of (\ref{12}). Using the abbreviations $l'_{a}=g'_{ab}l^{b}$,
   $n'_{a}=g'_{ab}n^{b}$ etc, one obtains after some rearrangement
   \begin{equation}\label{24}
   l'_{a}\tilde{r_{b}}+\tilde{r_{a}}l'_{b}-x'_{a}\tilde{s_{b}}
   -\tilde{s_{a}}x'_{b}=\beta(y'_{a}l_{b}+l_{a}y'_{b})
   \end{equation}
   where $\tilde{r_{a}}\equiv \alpha x_{a}+\beta
   y_{a}+2/3R'_{ab}x^{b}$ and $\tilde{s_{a}}=\alpha
   l_{a}+2/3R'_{ab}l^{b}$. Next, following the procedure used in the
   previous lemma, one expands $\tilde{r}, \tilde{s}$ and $l$ in
   terms of $l', n', x'$ and $y'$, having first used a rotation in the
   $x,y$ plane to eliminate the $x'$ term in the expansion for $l$.
   This simply changes the eigenbivector $\overset{+}{F}$ by multiplying it
   by $e^{i\theta}$,  $(\theta\in \R)$, and this (duality rotation) does not affect
   any other part of the calculation. On substituting these
   expressions into (\ref{24}) one finds, at $p$, that if $\beta\neq 0$,
   $l'_{a}\equiv g'_{ab}l^{b}=\kappa l_{a}$ $(\kappa \in \R$) and
   that $(\lambda g'_{ab}-2/3R'_{ab})l^{b}=\alpha l_{a}$, $(\lambda
   \in \R)$. Thus $l$ is a (g-null) eigenvector of $g'_{ab}$ and $R'_{ab}$, as
   required. If $\beta=0$ at $p$, one similarly finds that
   $l_{a}=\kappa'l'_{a}+\lambda'n'_{a}$, $(\kappa', \lambda'\in \R)$
   and that there exists $\kappa''\in \R$ such that $l$ and $x$ are
   eigenvectors of $(\kappa''g_{ab}-2/3R'_{ab})$ with eigenvalue $\alpha$. By
   repeating this last case with $F$ replaced by $\overset{*}{F}$ and using
   the second equation in (\ref{12}) one sees that (cf. lemma $3$) that $l,x$
   and $y$ are eigenvectors of $(\kappa''g_{ab}-2/3R'_{ab})$ with eigenvalue
   $\alpha$ at $p$.

   For lemma $3$ it is clear from lemmas $1(ii)$ and $2(ii)$ that if
   $B$ and $B'$ are any two members of the set of blades described
   in lemma 3, then each is an $\alpha$-eigenspace of a tensor of
   the form $(\tau g'_{ab}-2/3R'_{ab})$ for some $\tau\in \R$ which
   depends on the eigenspace. If these blades have a common non-zero
   member $v\in T_{p}M$ then, with $\tau'\in \R$, $(\tau
   g'_{ab}-2/3R'_{ab})v^{b}=\alpha v_{a}=(\tau'
   g'_{ab}-2/3R'_{ab})v^{b}$. It follows that $\tau =\tau'$ and
   lemma $3$ follows.

   In the next section, it will often be convenient to cast certain
   symmetric second order tensors into a canonical form based on
   their Segre type (Jordan canonical form) with respect to $g$ (see just before lemma 1). This, of course, can be
   done without knowledge of the latter theory by writing out the tensor in
   terms of symmetrised products of basis members. If required, details of this
   Segre classification can be found in \cite{4}.
   \section{The Main Theorem}
   The main theorem can now be stated.

   \begin{Theorem}
   Let $M$ be a space-time manifold on which two
   Lorentz metrics $g$ and $g'$ are defined with respective Levi-Civita
   connections $\nabla$ and $\nabla'$ and corresponding curvatures
   $R$ and $R'$. The geodesics associated with $\nabla$ and
   $\nabla'$ are taken to satisfy assumption A of section 1 and so $\nabla$ and $\nabla'$ are projectively related. Let $g$ be a non-flat vacuum metric.
   Then $\nabla$ and $\nabla'$ are necessarily equal on $M$ and so
   $g'$ is necessarily also a vacuum metric. If, in addition, the
   vacuum curvature tensor $R$ is such that, in the decompositions
   (\ref{10}) and (\ref{11}), any of the sets \ $\mb{I}$, $\mb{II}$, $\mb{D}$ and $\mb{III}$ is non empty then $g'=cg$ where $0<c\in \R$.
   Otherwise, the situation is as described in
   section 2 (since now $g$ and $g'$ are each vacuum metrics).
\end{Theorem}

$\mb{Proof}$ Each Petrov region will be considered in turn.

First consider the region $\mb{III}$ and let $p\in\mb{III}$. Then
one can choose a canonical null tetrad $l,n,x,y$ at $p$ such that
the vacuum curvature tensor $R$ takes the form \cite{12,13,14,4}
\begin{equation}\label{25}
R_{abcd}=a(V_{ab}M_{cd}+M_{ab}V_{cd}-\overset{*}{M}_{ab}\overset{*}{V}_{cd}
-\overset{*}{V}_{ab}\overset{*}{M}_{cd})
\end{equation}
where $V=\case{1}{\sqrt{2}}l\wedge x$, $\overset{*}{V}=-\case{1}{\sqrt{2}}l\wedge y$, $M=l\wedge n$,
$\overset{*}{M}=x\wedge y$ and $0\neq a\in\R$ and where $l$ is the (unique)
repeated principal null direction of $R$ at $p$. The only self-dual complex
eigenbivectors of $R$ are non-zero complex multiples of the complex
null bivector $V+\rmi \overset{*}{V}$ and the associated eigenvalue is zero.
Thus $R_{abcd}V^{cd}=R_{abcd}\overset{*}{V}{}^{cd}=0$ (recalling that all indices
are manipulated with the metric $g$). Thus, from lemma 2$(ii)$,
there exists $\delta\in\R$ such that $l,x$ and $y$ are eigenvectors
of $\delta g'_{ab}-\case{2}{3}R'_{ab}$ each with zero eigenvalue.
Thus, at $p$, $\delta g'_{ab}-\case{2}{3}R'_{ab}=bl_al_b$
($b\in\R$). Solving this expression for $R'_{ab}$ and substituting
into (\ref{13}) gives
\begin{equation}\label{26}
g'_{ae}R^e_{\ bcd} + g'_{be}R^e_{\ acd} =
\case{1}{2}b(g'_{ac}l_bl_d+g'_{bc}l_al_d-g'_{ad}l_bl_c-g'_{bd}l_al_c)
\end{equation}
Now (\ref{26}) when contracted with $x^cy^d$ and use is made of (\ref{25}) yields
$g'_{ae}\overset{*}{V}{}^{e}_{\ b}+g'_{be}\overset{*}{V}{}^{e}_{\ a}=0$. It follows \cite{15,4}
that the blade of $\overset{*}{V}$ is an eigenspace of $g'$ and hence that
\begin{equation}\label{27}
g'_{ab}=a_1g_{ab}+a_2l_al_b+a_3x_ax_b+a_4(l_ax_b+x_al_b)
\end{equation}
with $a_1,...,a_4\in\R$ and $a_1\neq0$ (since $g'$ is non-singular).
A contraction of (\ref{26}) with $n^ay^bn^cy^d$ and use of
(\ref{25}) and (\ref{27}) then gives
$2aa_4=-ba_1$.
However, a similar contraction with $n^an^bl^cn^d$ gives $2aa_4=ba_1$. Together these two results are consistent only if $aa_4=0$ and $ba_1=0$.
Thus $b=0$ and hence $g'_{ae}R^e_{\ bcd} + g'_{be}R^e_{\ acd}=0$. From this and the fact that the rank of the curvature tensor in its $6\times6$ form is four at each $p\in\mb{III}$ it
follows \cite{15,9,4} that $g'=a_1g$ at $p$. Thus $g$ and $g'$ are
conformally related at each $p\in\mb{III}$. Now write $g'=\phi g$ on
$\mathrm{int}\mb{III}$ for some (necessarily smooth) function
$\phi:\Int\mb{III}\rightarrow\R$ and substitute into (\ref{4}). For
any $p\in\Int\mb{III}$ choose $k\in T_pM$ such that $k^a\psi_a=0\neq
g_{ab}k^ak^b$ and contract (\ref{4}) with $k^ak^b$ to get
$\phi_{,a}=2\phi\psi_a$. A back substitution and contraction with
$k^b$ then reveals that $\psi=0$ at $p$ and hence on $\Int\mb{III}$.
Thus $\psi_{\,a}=0$ on $\Int\mb{III}$ and so $\phi$ is constant on
each component of $\Int\mb{III}$. [This last result could also be
found more generally from \cite{9,4}.]
In conclusion, $\nabla=\nabla'$ on $\Int\mb{III}$ and $g'=\phi g$ for constant $\phi$ on each component of $\Int\mb{III}$.
\\

Now let $p$ be a point in the region $\mb{D}$. At $p$, one may choose a
canonical null tetrad $l,n,x,y$ at $p$ so that $R$ takes the form
\cite{12,13,14,4}
\begin{equation}\label{28}
R_{abcd}=Re\{z(\overset{+}{V}_{ab}\overset{+}{U}_{cd}+\overset{+}{U}_{ab}\overset{+}{V}_{cd}
+\overset{+}{M}_{ab}\overset{+}{M}_{cd})\}
\end{equation}
where $\overset{+}{V}=V+\rmi \overset{*}{V}$, $\overset{+}{M}=M+\rmi \overset{*}{M}$ (with $V$ and $M$ as
before) and $\overset{+}{U}=U+\rmi \overset{*}{U}$, where $U=\case{1}{\sqrt{2}}n\wedge x$ and
$\overset{*}{U}=\case{1}{\sqrt{2}}n\wedge y$ and $0\neq z\in\C$ and where $l$ and $n$ span the
repeated principal null directions of $R$ at $p$. The complex self
dual eigenbivectors of $R$ are $\overset{+}{V}$ and $\overset{+}{U}$, each with eigenvalue
$z$ and $\overset{+}{M}$ with eigenvalue $-2z$. For this type, as in others to
follow, the cases when $z\in\R$ and when $z\notin\R$ are considered
separately.

If $z\in\R$ the null eigenbivectors $\overset{+}{V}$ and $\overset{+}{U}$
together with lemmas 2($ii$) and 3 show that there exists
$\delta\in\R$ such that $l$, $n$, $x$ and $y$ are each eigenvectors
of $\delta g'_{ab}-\case{2}{3}R'_{ab}$ with the same eigenvalue
$z$. Hence
\begin{equation}\label{29}
\delta g'_{ab}-\case{2}{3}R'_{ab}=zg_{ab}
\end{equation}
Then use of the non-null eigenvector $\overset{+}{M}$ and lemma 1($ii$) shows
that there exist $\delta', \delta''\in\R$ such that the blades of
$l\wedge n$ and $x\wedge y$ are eigenspaces of
$\delta'g'_{ab}-\case{2}{3}R'_{ab}$ and
$\delta''g'_{ab}-\case{2}{3}R'_{ab}$, respectively, with eigenvalue
$-2z$. Thus, by a rotation of the tetrad members $x$ and $y$ in the
blade of $x\wedge y$, if necessary, one gets for $b_1,...,b_5\in\R$
\begin{equation}\label{30}
\delta'g'_{ab}-\case{2}{3}R'_{ab}=-2zg_{ab}+b_1x_ax_b+b_2y_ay_b
\end{equation}
\begin{equation}\label{31}
\delta''g'_{ab}-\case{2}{3}R'_{ab}=-2zg_{ab}+b_3l_al_b+b_4n_an_b+b_5(l_an_b+n_bl_a)
\end{equation}
If $\delta'=\delta''$, (\ref{30}) and (\ref{31}), when subtracted, show that
$b_1=...=b_5=0$ and then (\ref{29}) and (\ref{30}), when subtracted, reveal that $g'\propto g$. If $\delta'\neq\delta''$ then a substitution of (\ref{29}) into
(\ref{30}) and (\ref{31}) gives
\begin{equation}\label{32}
(\delta'-\delta)g'_{ab}=-3zg_{ab}+b_1x_ax_b+b_2y_ay_b
\end{equation}
\begin{equation}\label{33}
(\delta''-\delta)g'_{ab}=-3zg_{ab}+b_3l_al_b+b_4n_an_b+b_5(l_an_b+n_bl_a)
\end{equation}
Thus $\delta'-\delta\neq 0\neq\delta''-\delta$ since $z\neq0$. Then,
on contracting (\ref{32}) and (\ref{33}) with $l^b$, one finds
first that $l'_a\propto l_a$ and then that $b_4=0$. A similar contraction with $n^b$ reveals that $b_3=0$. Thus
\begin{equation}\label{34}
g'_{ab}=cg_{ab}+d(l_an_b+n_bl_a)
\end{equation}
where $d\in\R$ and this covers all possibilities when $z\in\R$.

If $z\notin\R$,
lemma 2($i$) shows that $l$ and $n$ are eigenvectors of $g'_{ab}$
and of $R'_{ab}$ at $p$. Since they are {\it $g$-null}, the eigenvalues of $l$ and $n$ are equal in each case (see, e.g. \cite{4}). Using this
information in lemma 1($i$) (equation (\ref{14})),
one sees that $x$ and $y$ are eigenvectors of $g'$ with the same eigenvalue. It has thus been shown that the
blades of $l\wedge n$ and $x\wedge y$ are each eigenspaces of $g'$
and hence that (\ref{34}) holds also in this case.

Thus (\ref{34})
holds at each point of the region $\mb{D}$. It now follows that if
$p\in\Int\mb{D}$, there exists an open neighbourhood $V\subset\Int\mb{D}$ of $p$ such
that $l$ and $n$ are smooth vector fields on $V$ because the Petrov type is constant there \cite{16} and that (\ref{34}) holds on $V$. To see
that $c$ and $d$ are smooth functions on $V$ one contracts
(\ref{34}) with the (smooth) tensors $l^b$ and $g^{ab}$ to see that
$c+d$ and $4c+2d$ are each smooth on $V$ and the result follows
(and, in addition, it follows from the non-degeneracy of $g'$ that
$c$ and $c+d$ are nowhere zero on $V$). Now substitute
(\ref{34}) into (\ref{4}) and contract successively with $l^al^b$,
$n^an^b$, $x^ax^b$ and $y^ay^b$ to get
$\psi_al^a=\psi_an^a=\psi_ax^a=\psi_ay^a=0$ and so the 1-form $\psi$
is zero on $V$ and hence on $\Int\mb{D}$. Thus, from (\ref{3}), the
connections $\nabla$ and $\nabla'$ and hence their associated type
($1,3$) curvature tensors are equal on $\Int\mb{D}$. It follows
\cite{9,4} that $g'$ and $g$ are related by a {\it constant}
conformal factor on each component of $\Int\mb{D}$.
\\

Now let $p$ be a point in the region $\mb{II}$. At $p$, one may
choose a null tetrad $l,n,x,y$ at $p$ so that $R$ takes the form
(using a previously established notation; \cite{12,13,14,4})
\begin{equation}\label{35}
R_{abcd}=Re\{z_1\overset{+}{V}_{ab}\overset{+}{V}_{cd}
+z_2(\overset{+}{V}_{ab}\overset{+}{U}_{cd}
+\overset{+}{U}_{ab}\overset{+}{V}_{cd}+\overset{+}{M}_{ab}\overset{+}{M}_{cd})\}
\end{equation}
with $0\neq z_1\in\R$ and $0\neq z_2\in\C$. Here $l$ spans the unique repeated principal null direction of $R$
at $p$. The complex self dual eigenbivectors of $R$ are the null bivector $\overset{+}{V}$ with
eigenvalue $z_2$ and the non-null bivector $\overset{+}{M}$ with eigenvalue
$-2z_2$. Again, the cases when $z_2\in\R$ and when $z_2\notin\R$ are
considered separately. So consider first the case when $z_2\in\R$.
Then lemma 2($ii$) shows that there exists $\delta\in\R$ such that
$l$, $x$ and $y$ are eigenvectors of $\delta
g'_{ab}-\case{2}{3}R'_{ab}$ with the same eigenvalue $z_2$ and so,
at $p$
\begin{equation}\label{36}
\delta g'_{ab}-\case{2}{3}R'_{ab}=z_2g_{ab}+\sigma l_al_b
\end{equation}
for $\sigma\in\R$. Next, lemma 1($ii$) confirms the existence of
$\delta', \delta''\in\R$ such that the blades of $l\wedge n$ and
$x\wedge y$ are eigenspaces of $\delta'g'_{ab}-\case{2}{3}R'_{ab}$
and $\delta''g'_{ab}-\case{2}{3}R'_{ab}$, respectively, with
eigenvalue $-2z_2$ and so (recalling the completeness relation
$g_{ab}=l_an_b+n_al_b+x_ax_b+y_ay_b$)
\begin{equation}\label{37}
\delta'g'_{ab}-\case{2}{3}R'_{ab}=-2z_2g_{ab}+a_1x_ax_b+a_2y_ay_b+a_3(x_ay_b+y_ax_b)
\end{equation}
\begin{equation}\label{38}
\delta''g'_{ab}-\case{2}{3}R'_{ab}=-2z_2g_{ab}+a_4l_al_b+a_5n_an_b+a_6(l_an_b+n_bl_a)
\end{equation}
where $a_1,...,a_6\in\R$. If
$\delta'=\delta''$, (\ref{37}) and (\ref{38}) show that
$a_1=...=a_6=0$ and then (\ref{36}) and (\ref{37}), on subtraction, reveal that $g'_{ab}$ is a linear combination of $g_{ab}$ and $l_al_b$.
Otherwise, if $\delta'\neq\delta''$, then on substituting (\ref{36})
into (\ref{37}) and (\ref{38}), one obtains expressions for
$(\delta'-\delta)g'$ and $(\delta''-\delta)g'$ from which it is
clear that $\delta'-\delta\neq 0\neq\delta''-\delta$ since
$z_2\neq0$ and that the blade of $x\wedge y$ is an eigenspace of
$g'$ and that $l$ is also an eigenvector of
$g'$. Thus at $p$
\begin{equation}\label{39}
g'_{ab}=c_1g_{ab}+c_2(l_an_b+n_bl_a)+c_3l_al_b
\end{equation}
where $c_1(\neq0), c_2, c_3\in\R$. Thus (\ref{39}) holds whether $\delta'=\delta''$ or not.
Substituting (\ref{39}) and (\ref{36}) into (\ref{13}), recalling
(\ref{35}), contractions with $n^ax^bn^cx^d$ and
$n^ay^bn^cy^d$ show, respectively, that
$c_1\sigma=-2c_2z_1$ and $c_1\sigma=2c_2z_1$. These relations imply
$c_2z_1=0=c_1\sigma$. Now $z_1\neq0\neq c_1$ so $c_2=0=\sigma$. Finally, a contraction with $n^an^bl^cn^d$ then shows $0=z_2c_3$, hence $c_3=0$. Thus $g'\propto g$ at $p$.

If $z_2\notin\R$, lemmas 2($i$) and 1($i$) can be applied, respectively, to
the bivectors $\overset{+}{V}$ and $\overset{+}{M}$. The first of these applications
reveals that $l$ is an eigenvector of $g'_{ab}$ and of $R'_{ab}$ at
$p$. In the second application (using (\ref{14}) and (\ref{15}))
one can then incorporate the conditions that $l^a$ is
an eigenvector of $g'_{ab}$ and of $R'_{ab}$ (so that $l'\propto l$
and $\tilde{R}l\propto l$). Thus one finds (for $b_1,...,b_6\in\R$)
\begin{equation}\label{40}
\begin{array}{ll}
  \beta l=b_1l' & \alpha l+\case{2}{3}R'l=b_5l' \\
  \beta n=b_1n'+b_2x'+b_3y' & \alpha n+\case{2}{3}R'n=b_5n'+b_3x'-b_2y' \\
  \beta x=b_4x'+b_2l' & \alpha x+\case{2}{3}R'x=b_6x'+b_3l' \\
  \beta y=b_4y'+b_3l' & \alpha y+\case{2}{3}R'y=b_6y'-b_2l'
\end{array}
\end{equation}
The left hand set of equations can be inverted to give (with
$c_1,...,c_4\in\R$) \begin{equation}\label{41}
\begin{array}{l}
  \beta l'=c_1l  \\
  \beta n'=c_1n+c_2x+c_3y+c^{-1}_4(c_2^2+c^3_2)l  \\
  \beta x'=c_4x+c_2l \\
  \beta y'=c_4y+c_3l
\end{array}
\end{equation}
Equation (\ref{41}) provides algebraic information on $g'$ (since $\beta l'_a=\beta g'_{ab}l^b$ etc) and one finds
\begin{equation}\label{42}
\fl g'_{ab}=d_1(l_an_b+n_al_b)+d_2(x_ax_b+y_ay_b)+d_3(l_ax_b+x_al_b)
+d_4(l_ay_b+y_al_b)+d^{-1}_2(d_3^2+d_4^2)l_al_b
\end{equation}
where $d_1,...,d_4\in\R$.
The second column of equations in (\ref{40}), together with (\ref{41}) reveal similar algebraic information on $R'_{ab}$ to that on $g'_{ab}$ with the result that one can write (with $d'_1,...,d'_5\in\R$
\begin{equation}\label{43}
\fl R'_{ab}=d'_1(l_an_b+n_al_b)+d'_2(x_ax_b+y_ay_b)+d'_3(l_ax_b+x_al_b)
+d'_4(l_ay_b+y_al_b)+d'_5l_al_b
\end{equation}
Substituting (\ref{42}) and (\ref{43}) into (\ref{13}), recalling
(\ref{35}), contractions with $l^ax^bn^cy^d$, $y^ay^bn^cx^d$ and
$x^ax^bn^cy^d$ show that $d_1=d_2$, $d_4=0$ and $d_3=0$ in (\ref{42}).
Thus $g'\propto g$ at $p$ and so $g$ and $g'$ are conformally related on $\mb{II}$. As in the previous cases, one can now use
(\ref{4}) or, more generally \cite{9,4}, to show that $g$ and $g'$ are conformally related with a constant conformal factor on each component of $\Int\mb{II}$. It follows that $\nabla=\nabla'$ on $\Int\mb{II}$.
\\

Now let $p$ be a point in the region $\mb{I}$. At $p$, the curvature
tensor $R$ admits three independent complex self dual eigenbivectors which,
in terms of a pseudo-orthonormal (canonical Petrov) tetrad $u, x, y, z$,
may be taken in the form $u\wedge z+\rmi x\wedge y$, $u\wedge y+\rmi z\wedge x$ and
$u\wedge x+\rmi y\wedge z$. Their respective eigenvalues are the complex
numbers $z_1$, $z_2$ and $z_3$ and they are {\it distinct}. The trace-free condition
on $R$ means that $z_1+z_2+z_3=0$. The trace-free condition shows that three cases may be distinguished and these turn out to be convenient for the present purpose. Case 1 is when each eigenvalue is real, case 2 when no eigenvalue is real and case 3 when two eigenvalues are not real and one is real.

In case 1, $z_1,z_2,z_3\in\R$. Then lemma 1($ii$) applied to the above eigenbivectors shows the existence of $a_i\in\R$ ($1\leq i\leq6$) such that the tensors $a_ig'_{ab}-\case{2}{3}R'_{ab}$ admit the blades of $u\wedge z$, $x\wedge y$, $u\wedge y$, $z\wedge x$, $u\wedge x$ and $y\wedge z$, respectively, as eigenspaces with respective eigenvalues $z_1, z_1, z_2, z_2, z_3, z_3$ ($z_3=-(z_1+z_2)$). So writing out (\ref{16}) for the complex eigenbivectors given in the previous paragraph, one finds in an obvious shorthand notation with indices suppressed
\begin{equation}\label{44}
\begin{array}{l} z_1 u+\case{2}{3}\tilde{R}u=a_1u' \\
z_1 z+\case{2}{3}\tilde{R}z=a_1z' \\
z_1 x+\case{2}{3}\tilde{R}x=b_1x' \\
z_1 y+\case{2}{3}\tilde{R}y=b_1y'\\
\end{array}\ \ \ \
\begin{array}{l} z_2 u+\case{2}{3}\tilde{R}u=a_2u' \\
z_2 y+\case{2}{3}\tilde{R}y=a_2y'\\
z_2 z+\case{2}{3}\tilde{R}z=b_2z' \\
z_2 x+\case{2}{3}\tilde{R}x=b_2x' \\
\end{array}\ \ \ \
\begin{array}{l} z_3u+\case{2}{3}\tilde{R}u=a_3u' \\
z_3x+\case{2}{3}\tilde{R}x=a_3x' \\
z_3z+\case{2}{3}\tilde{R}z=b_3z' \\
z_3y+\case{2}{3}\tilde{R}y=b_3y'\\
\end{array}\end{equation}
Subtracting the $"u"$ equations in columns one and two shows that {$(z_1-z_2)u_a=(a_1-a_2)g'_{ab}u^b$}. Thus, since $z_1\neq z_2$, $a_1\neq a_2$ and one concludes that $u$ is an eigenvector of $g'$ with eigenvalue $\case{z_1-z_2}{a_1-a_2}$. Similar subtractions of the $"z"$, $"x"$ and $"y"$ equations from columns one and two show that $z$, $x$ and $y$ are also eigenvectors of $g'$. Thus, at $p$, $g'_{ab}=-\gamma_0u_au_b+\gamma_1x_ax_b+\gamma_2y_ay_b+\gamma_3z_az_b$, $\gamma_0,...,\gamma_3\in\R$. Repeating the procedure for columns two and three and for columns one and three and equating the various expressions for the eigenvalues of $u, z, x$ and $y$ gives the four groups of equations
\begin{equation}\label{45}
\begin{array}{l}
\begin{array}{lll}
(z_1-z_2)&=(a_1-a_2)\gamma_0 \hspace{1cm} (z_1-z_2)&=(a_1-b_2)\gamma_3 \\
(z_1-z_3)&=(a_1-a_3)\gamma_0 \hspace{1cm} (z_1-z_3)&=(a_1-b_3)\gamma_3 \\
(z_2-z_3)&=(a_2-a_3)\gamma_0 \hspace{1cm} (z_2-z_3)&=(b_2-b_3)\gamma_3 \\
\end{array} \\  \\
\begin{array}{lll}
(z_1-z_2)&=(b_1-b_2)\gamma_1 \hspace{1cm} (z_1-z_2)&=(b_1-a_2)\gamma_2 \\
(z_1-z_3)&=(b_1-a_3)\gamma_1 \hspace{1cm} (z_1-z_3)&=(b_1-b_3)\gamma_2 \\
(z_2-z_3)&=(b_2-a_3)\gamma_1 \hspace{1cm} (z_2-z_3)&=(a_2-b_3)\gamma_2 \\
\end{array}
\end{array}
\end{equation}
Now all the bracketed terms in the above equations are non-zero and so, eliminating the $\gamma$ terms, one finds
\[ \frac{z_1-z_2}{a_1-a_2}=\frac{z_1-z_3}{a_1-a_3}=\frac{z_2-z_3}{a_2-a_3} \hspace{1cm}
\frac{z_1-z_2}{a_1-b_2}=\frac{z_1-z_3}{a_1-b_3}=\frac{z_2-z_3}{b_2-b_3}
\]
\begin{equation}\label{46}\end{equation}
\[ \frac{z_1-z_2}{b_1-b_2}=\frac{z_1-z_3}{b_1-a_3}=\frac{z_2-z_3}{b_2-a_3} \hspace{1cm} \frac{z_1-z_2}{b_1-a_2}=\frac{z_1-z_3}{b_1-b_3}=\frac{z_2-z_3}{a_2-b_3}
\]
The first equation in each of the collection (\ref{46}) when subtracted from each other in an appropriate way yields $(z_1-z_2)(a_3-b_3)=(z_1-z_3)(a_2-b_2)=(z_1-z_3)(b_2-a_2)$ and so $a_2=b_2$ and $a_3=b_3$. Similarly, using the second equation of each of (\ref{46}), one finds, in addition, $a_1=b_1$. Then (\ref{45}), together with the knowledge that each of the bracketed quantities in (\ref{46}) is non-zero, shows that $\gamma_1=\gamma_2=\gamma_3=\gamma_0$ and so $g'\propto g$ at $p$.

In case 2, when none of $z_1, z_2$ and $z_3$ is real, one appeals to lemma 1($i$) which for the bivector $u\wedge z+\rmi x\wedge y$ is written out in (\ref{14}) and (\ref{15}) with $z_1=\alpha+\rmi \beta$ and $\beta\neq0$. Now consider the corresponding results for the bivectors $u\wedge y+\rmi z\wedge x$ and $u\wedge x+\rmi y\wedge z$. Equation (\ref{14}) shows that $u$ is a linear combination of $u', x'$ and $y'$. The corresponding equations for the other two bivectors reveal, on the other hand, that $u$ is a linear combination of $u', x'$ and $z'$ and of $u', y'$ and $z'$, respectively. It follows that, in (\ref{14}), $b_{02}=b_{03}=0$, and with similar vanishing coefficients in the other two sets of equations. Thus $u'\propto u$ and so $u$ is an eigenvector of $g'$. Applying similar arguments to $x$, $y$ and $z$ shows that they are also eigenvectors of $g'$. An inspection of 
any two of the three sets of equations of the form (\ref{14})
then reveals that the eigenvalues of $g'$ associated with $u, x, y$ and $z$ are equal and so $g'\propto g$ at $p$.

In case 3, suppose that the eigenvalue $z_1=\alpha_1$, corresponding to $u\wedge z+\rmi x\wedge y$ is real. The tracefree condition then shows that $u\wedge y+\rmi z\wedge x$ and $u\wedge x+\rmi y\wedge z$ have eigenvalues $\alpha_2+\rmi \beta_2$ and $\alpha_3-\rmi \beta_2$ ($\alpha_2, \alpha_3, \beta_2\in\R,\ \beta_2\neq0$). So lemma 1($ii$) applies to the first of these and lemma 1($i$) applies to the other two. In the latter case, an inspection of the two sets of equations corresponding to (\ref{14}) similar to that in case 2 reveals that, at $p$, $u$ and $z$ are linear combinations of $u'$ and $z'$ only and that $x$ and $y$ are linear combinations of $x'$ and $y'$ only. After removing the zero coefficients and making the obvious identifications, one has
\begin{equation}\label{47}
\begin{array}{lll}
-\beta_2u=cu'-d_1z' & \hspace{1cm} & -\beta_2x=cx'+d_2y' \\
-\beta_2z=cz'+d_1u' & \hspace{1cm} & -\beta_2y=cy'+d_2x'
\end{array}
\end{equation}
where $c,d_1,d_2\in\R$. Together with the information (from lemma 1(ii)) for the eigenbivector with real eigenvalue, the remaining information from the eigenbivectors with non-real eigenvalues is, using a previous abridged notation, respectively,
\begin{equation}\label{48}
\fl\begin{array}{l}
\alpha_1u+\case{2}{3}\tilde{R}u=au' \\
\alpha_1z+\case{2}{3}\tilde{R}z=ay' \\
\alpha_1x+\case{2}{3}\tilde{R}x=bx' \\
\alpha_1y+\case{2}{3}\tilde{R}y=bz' \\
\end{array} \hspace{5mm}
\begin{array}{l}
\alpha_2u+\case{2}{3}\tilde{R}u=d_3u'+d_2z' \\
\alpha_2y+\case{2}{3}\tilde{R}y=d_3y'+d_1x' \\
\alpha_2x+\case{2}{3}\tilde{R}x=d_4x'+d_1y' \\
\alpha_2z+\case{2}{3}\tilde{R}z=d_4z'-d_2u' \\
\end{array} \hspace{5mm}
\begin{array}{l}
\alpha_3u+\case{2}{3}\tilde{R}u=d_5u'+d_2z' \\
\alpha_3x+\case{2}{3}\tilde{R}x=d_5x'+d_1y' \\
\alpha_3z+\case{2}{3}\tilde{R}z=d_6z'-d_2u' \\
\alpha_3y+\case{2}{3}\tilde{R}y=d_6y'+d_1x' \\
\end{array}
\end{equation}
where $a,b,d_3,...,d_6\in\R$. If one subtracts, respectively the ``$u$'', ``$z$'', ``$x$'' and ``$y$'' equations in the second and third sets of equations in (\ref{48}), one finds
\begin{equation}\label{49}
\begin{array}{l}
(\alpha_3-\alpha_2)u=(d_5-d_3)u'  \hspace{1cm} (\alpha_3-\alpha_2)x=(d_5-d_4)x'\\
(\alpha_3-\alpha_2)z=(d_6-d_4)z'  \hspace{1cm} (\alpha_3-\alpha_2)y=(d_6-d_3)y'\\
\end{array}
\end{equation}
If $\alpha_2=\alpha_3$ then (\ref{49}) gives $d_3=d_5=d_4=d_6$ and then, similarly, from
the first and second sets of equations (\ref{49}), one finds
\begin{equation}\label{50}
\begin{array}{l}
(\alpha_2-\alpha_1)u=(d_3-a)u'+d_2z' \hspace{1cm} (\alpha_2-\alpha_1)x=(d_3-b)x'+d_1y'\\
(\alpha_2-\alpha_1)z=(d_3-a)z'-d_2u' \hspace{1cm} (\alpha_2-\alpha_1)y=(d_3-b)y'+d_1x'\\
\end{array}
\end{equation}
Now from the $z'$ coefficient in the ``$u$'' equations in (\ref{47}) and (\ref{50}) one finds $(\alpha_2-\alpha_1)d_1=\beta_2d_2$, while from the $y'$ coefficient in the ``$x$'' equations one finds $-(\alpha_2-\alpha_1)d_2=\beta_2d_1$. These two results together imply that $((\alpha_2-\alpha_1)^2+\beta_2^2)d_2=0$, which, since $\beta_2\neq0$, means $d_2=0$ and hence $d_1=0$. It now follows from (\ref{47}) that each of $u,z,x$ and $y$ is an eigenvector of $g'$, with equal eigenvalues.
If $\alpha_2\neq\alpha_3$, (\ref{49}) immediately implies that each of $u,z,x$ and $y$ is an eigenvector of $g'$, with (\ref{47}) further implying that all eigenvalues are equal and that $d_1=d_2=0$.
So, in either case $u, z, x$ and $y$ are eigenvectors of $g'$ with equal eigenvalue and hence $g'\propto g$ at $p$.

Thus, on the open region $\mb{I}$ of $M$, $g'=\phi g$ for some necessarily smooth function $\phi:\mb{I}\rightarrow\R$. An argument identical to that in the other cases using (\ref{4}) or, more generally \cite{9,4}, then shows that, on $\mb{I}$, $\phi_{,a}=0$, $\psi=0$, $\nabla'=\nabla$ and $\phi$ is constant, on each component of $\mb{I}$.
\\

Finally, consider the region $\mb{N}$ and let $p\in\mb{N}$. Then
a null tetrad $l,n,x,y$ can be chosen at $p$ so that $l$ is the
unique repeated null direction of $R$ and so that, with $V$ as before and $0\neq c\in\R$,
\begin{equation}\label{51}
R_{abcd}=c(V_{ab}V_{cd}-\overset{*}{V}_{ab}\overset{*}{V}_{cd})\
\end{equation}
The self dual eigenbivectors of $R$ are $\overset{+}{V}$ and $\overset{+}{M}$ and in each
case the eigenvalue is zero. Thus $R$ has real eigenbivectors $V,
\overset{*}{V}, M$ and $\overset{*}{M}$, each with zero eigenvalue and, because of their
blade interactions, lemma 3 shows there exists $\delta\in\R$ such
that $l,n,x$ and $y$ are eigenvectors with zero eigenvalue of $\delta
g'_{ab}-\case{2}{3}R'_{ab}$. Thus $R'_{ab}=
\case{3\delta}{2}g'_{ab}$ and so, from (\ref{13}), $g'_{ae}R^e_{\
bcd}+g'_{be}R^e_{\ acd}=0$. It follows from \cite{15,9,4} that, at $p$,
$g'_{ab}=ag_{ab}+bl_al_b$ ($a,b\in\R$, $a\neq0$). Now if
$p\in\Int\mb{N}$ one can regard the last equation as holding on some
neighbourhood $V\subset\Int\mb{N}$ of $p$ with $a$ and $b$ real
valued functions on $V$. Because of the constancy of the Petrov type
on $\Int\mb{N}$ one can choose $V$ and then choose $l$ at each point
of $V$ so that it is smooth on $V$ \cite{16}. Then, it is easily checked that, with
this choice of $l$, the functions $a$ and $b$ are smooth. Now substitute
this relation into (\ref{4}) and contract at $p\in V$ with $k^ak^b$
where $k\in T_pM$ satisfies $l_ak^a=\psi_ak^a=0\neq g_{ab}k^ak^b$.
One finds, since $a$ is nowhere zero on $V$, that $a_{,a}=2a\psi_a$
and a back substitution and a contraction with $l^a$ show that
$l'_c\psi_b+(l^a\psi_a)g'_{bc}=0$. Thus $l_a\psi^a=0$ and then
$l'_a\psi_b=0$ at $p$. It follows that $\psi$ vanishes at $p$ and
hence on $\Int\mb{N}$ and so, from (\ref{3}), $\nabla=\nabla'$ on
$\Int\mb{N}$. Thus $g$ and $g'$ are each vacuum metrics and the
analysis given in section 2 completes this case.
\\

In conclusion, it has been shown that, in the decomposition
(\ref{11}), $\psi=0$ (and hence $\nabla'=\nabla$) on the open dense
subset $M\setminus E$ of $M$. Since $\psi$ is smooth on $M$,
$\psi\equiv0$ on $M$ and so $\nabla'=\nabla$ on $M$. It follows that
{\it $g'$ is a vacuum metric on $M$}. Further, in the decomposition
(\ref{10}), if any of the sets $\mb{I}$, $\mb{II}$, $\mb{D}$ or
$\mb{III}$ is non-empty, then, as mentioned in section 2, the
curvature tensor $R$ (or $R'$ since now $R'=R$) has rank at least 4.
Since $\nabla'=\nabla$ on $M$, it follows from holonomy theory that the common infinitesimal holonomy and holonomy algebras are of dimension $\geq4$. Thus
\cite{3,4} (see section 2) {\it $g$ and $g'$ are related on $M$
by a constant conformal factor} which must be positive to preserve
the Lorentz signature. If the regions $\mb{I}$, $\mb{II}$, $\mb{D}$
and $\mb{III}$ are each empty, the situation is as described in
section 2. This completes the proof of the theorem. $\square$

   \section{Remarks and Examples}
In this section some brief remarks are made regarding the main
theorem. This theorem, based only on assumption $A$ of section 1,
can be restated in the following way. Given the collection of
subsets $G_p\subset T_pM$ for each $p\in M$, as in section 1, and a
collection $C$ of curves in $M$ such that for any $p\in M$ and $v\in
T_pM$ there exists $c\in C$ starting from $p$ and whose tangent
vector at $p$ equals $v$, then there is at most one Levi-Civita
connection for which each member of $C$ is geodesic and whose Ricci
tensor is zero. If one exists then, under the highly non-restrictive
conditions stated in the main theorem, its compatible metric is
uniquely determined up to a constant conformal factor (or,
informally, "units of measurements"). Thus, in this case, the null
cone and collection of affine parameters along the geodesics are
determined. The first result given in section 2 of this paper (a weakened version
of the main theorem), when {\it both metrics $g$ and $g'$ were
assumed to be vacuum metrics}, is a reflection of the fact that a
vacuum metric cannot admit a {\it proper} projective vector field
\cite{17} and this latter result is easily derived from this
weakened version of the main theorem. To see this, briefly, let $X$
be a vector field on a vacuum space-time $(M,g)$ with local flows
denoted by $\phi_t$ (for more details, see e.g. \cite{4}). If a certain
$\phi_t$ has domain some open subset $U\subset M$, it is a
diffeomorphism $\phi_t:U\rightarrow\phi_t(U)$. Then $\phi_t$ is
called {\it projective} if whenever a path $c:I\rightarrow U$, where
$I$ is some open interval of $\R$, is (part of) an unparameterised
geodesic in $U$ then $\phi_t\circ c$ is (part of) an unparameterised
geodesic in $\phi_t(U)$. The map $\phi_t$ is called {\it affine} if
it is projective and, in addition, it preserves affine parameters
(in an obvious sense). Then the vector field $X$ is {\it projective}
if each of its local flows is projective and {\it affine} if each of
its local flows is affine. If $X$ is projective but not affine, it
is called {\it proper projective}. So if $X$ is projective, it can
be seen that $g$ and the pullback $\phi^*_tg$ of $g$ are metrics on
$U$ whose Levi-Civita connections are projectively related. But if
$g$ is a vacuum metric on $M$, $g$ and $\phi^*_tg$ restrict to
vacuum metrics on $U$ and the theorem shows that their Levi-Civita
connections are equal. Thus $\phi_t$ is affine for each $\phi_t$ and
so $X$ is affine and not proper projective.

The procedure of the previous paragraph, when applied to a space-time metric which admits a {\it proper} projective vector field, can be used to show that the totality of unparameterised geodesics does not determine the metric up to a constant factor since, for some local flow $\phi_t$ of this vector field, $\phi_t$ will preserve unparameterised geodesics but $g$ and $\phi^*_tg$ will differ by more than a constant conformal factor. For general techniques and examples see \cite{4,17,18}.
It is perhaps constructive, theoretically, to give some examples of situations where a collection of paths on $M$ could, simultaneously, be the collection of geodesics of the Levi-Civita connection of a space-time metric on $M$ and of a non-metric connection on $M$ (either symmetric or not).

   Let $\nabla$ be the Levi-Civita connection of some metric on $M$ with Christoffel symbols $\Gamma^{a}_{bc}$. Then, provided $M$ admits a global non-closed 1-form $\psi$, one can easily construct another symmetric connection $\nabla'$ on $M$ whose unparametrised geodesics are the same as those of $\nabla$, but which is not metric, by building it from the Christoffel symbols $\Gamma'^{a}_{bc}$ in each coordinate domain of $M$ according to the formula
   \begin{equation}\label{52}
   \Gamma'^{a}_{bc}=\Gamma^{a}_{bc}+\delta^{a}_{b}\psi_{c}+\delta^{a}_{c}\psi_{b}
   \end{equation}
   Then $\nabla'$ is not metric because, otherwise, the 1-form $\psi$ would be closed. [In fact one can also construct an example of such a symmetric {\it non-metric} connection $\nabla'$ in a similar way to this but with $\psi$ a {\it closed} 1-form.
   To see this, consider the following example (see and cf. \cite{19,20,21}). Let $U$ be the open subset of $R^{4}$ given by $x^{0}>0$ and let $\varphi:U\rightarrow\R$ be given by $\varphi(x^{0},x^{1},x^{2},x^{3})$=log $x^{0}$. Then with $\varphi_{a}\equiv\varphi,_{a}$, one has $\varphi_{a;b}=-\varphi_{a}\varphi_{b}$. Define a metric $g$ on the (global) chart $U$ by $g=e^{-2\varphi}\eta$, where $\eta$ is the Minkowski metric on $U$, and let $\Gamma^{a}_{bc}$ be its associated Levi-Civita connection coefficients on $U$ (which then define a symmetric, metric connection $\nabla$ on $U$). Then, with a {\it closed} 1-form $\psi$ defined on $U$ by the components $\psi_{a}=(1-e^{\varphi})^{-1}\varphi_{a}$, define another connection $\nabla'$ on $U$ by (\ref{52}).
   Then $\nabla'$ is the desired connection on $U$ because it is known that $\nabla'$ is ${\it not}$ a metric connection \cite{19,20}]. Thus, knowledge of the total (unparametrised) geodesic
   structure of space-time does not determine whether the connection is metric or not.

   Now consider the metric $g$ given on $\R^{4}$ by
   \begin{equation} \label{53}
   ds^{2}=-dt^{2}+dx^{2}+q_{\alpha\beta}dx^{\alpha}dx^{\beta}
   \end{equation}
   where $q_{\alpha\beta}$ is a 2-dimensional positive definite metric on the $y,z$ plane which is not flat. This metric has holonomy group of the type $R_{4}$ (for details see \cite{4,10}).
   It admits a global spacelike (simple) bivector $F$ which is covariantly constant, and independent covariantly constant global (smooth) null vector fields $l$ and $n$ such that $F_{ab}l^{a}=F_{ab}n^{b}=0$ (and $l^an_a=1$).
   Now let $\nabla$ be the Levi-Civita connection arising from $g$ with Christoffel symbols $\Gamma^{a}_{bc}$ and define a connection $\nabla'$ on $\R^{4}$ by the Christoffel symbols $\Gamma'^{a}_{bc}=\Gamma^{a}_{bc}+l^{a}F_{bc}$. Then $\nabla$ and $\nabla'$ are easily seen to have the same affinely parameterised geodesics but $\nabla'$ is {\it neither symmetric nor metric}. That it is not symmetric is clear. To see, briefly, why it is not metric, let $P^{a}_{bc}=\Gamma'^{a}_{bc}-\Gamma^{a}_{bc}=l^{a}F_{bc}$. Then (see appendix) the Riemann tensors associated with $\nabla$ and $\nabla'$, with components $R'^a{}_{bcd}$ and $R^a{}_{bcd}$, are equal. Suppose $\nabla'$ is metric with compatible (smooth) metric $h$,
   and so $\nabla'h=0$. Then, even though $\nabla'$ is not symmetric, the (generalised) Ricci identity still yields $h_{ae}R'^e_{\ bcd}+h_{be}R'^e_{\ acd}=0$ \cite{7}. So the equality of the curvature tensors gives
   \begin{equation}\label{54}
   h_{ae}R^e{}_{bcd}+h_{be}R^e{}_{acd}=0
   \end{equation}
   From this, it follows \cite{4,9,15} that $h$ may be written in terms of $g$, $l$ and $n$ as $h_{ab}=\mu g_{ab}+ \varrho\, l_{a}l_{b}+\sigma n_{a}n_{b}+2\varsigma l_{(a}n_{b)}$ for appropriate real valued functions $\mu,\varrho,\,\sigma$ and $\varsigma$ on $R^{4}$ which are easily shown to be smooth and where the indices on the vector fields $l$ and $n$ are lowered with the original metric $g$. One then writes the difference between $\nabla h$ and $\nabla' h$ in terms of $P$. Since $\nabla' h=0$, this gives an expression for $\nabla h$ which can be equated to the $\nabla$-covariant derivative of the above expression for $h$. On contracting this expression successively with $l^{a}l^{b}$, $n^{a}n^{b}$ and $l^{a}n^{b}$ one finds that $\varrho$, $\sigma$ and $\mu+\varsigma$ are constant.
   A back substitution of this information into the previous expression
   and contractions with $l^a$ and $n^a$ show that $\sigma=\mu+\varrho=0$ and hence the contradiction that $h$ is degenerate.
   Thus, although $\nabla$ and $\nabla'$ have the same ${\it affinely parametrised}$ geodesics (and $\nabla$ is metric), $\nabla'$ is ${\it not}$ metric. [An alternative proof of the non-metric nature of $\nabla'$ can be obtained by computing $R'^a{}_{bcd|e}$, where a vertical stroke denotes a covariant derivative with respect to $\nabla'$ and then showing that, with the above expression for $h$, the condition $h_{ea}R^e{}_{bcd|f}+h_{eb}R^e{}_{acd|f}=0$, which is a necessary condition for $\nabla'$ to be metric, with compatible metric $h$, requires that $\sigma=\mu+\varsigma=0$. But this again leads to the contradiction that $h$ is degenerate.]
   \section{Appendix}

   To facilitate calculations such as some of those above, the following result, which very slightly, but importantly for a result in the last section, generalises an earlier one of Rosen \cite{22}, is given.

   Let $\nabla$ and $\nabla'$ be (not necessarily symmetric or metric) connections on a manifold $M$ such that, in some (any) coordinate domain $U$, their Christoffel symbols are $\Gamma^{a}_{bc}$ and $\Gamma'^{a}_{bc}$ and their curvature tensors
   $R^a{}_{bcd}$ and $R'^a{}_{bcd}$, respectively. Let $P^{a}_{bc}=\Gamma'^{a}_{bc}-\Gamma^{a}_{bc}$ and let a semi-colon denote a covariant derivative with respect to $\nabla$. Then if $H^a{}_{bcd}$ is defined as a tensor constructed in the usual manner from the curvature tensor, but with Christoffel symbols replaced by the tensor $P$ and partial derivatives replaced by $\nabla$ covariant derivatives, then one finds after some calculation that
   \begin{equation}\label{55}
   H^a{}_{bcd}\equiv P^{a}_{db;c}-P^{a}_{cb;d}+P^{e}_{db}P^{a}_{ce}-P^{e}_{cb}P^{a}_{de}
   =R'^a{}_{bcd}-R^a{}_{bcd}+P^{a}_{eb}Q^{e}_{dc}
   \end{equation}
   where $Q^{a}_{bc}\equiv\Gamma^{a}_{bc}-\Gamma^{a}_{cb}$ are the components of the
   ${\it torsion}$ tensor of $\nabla$. It is noted that, in (\ref{55}), the covariant derivatives on the left hand side and the torsion $Q$ on the right hand side are with respect to $\nabla$. When $\nabla$ is symmetric, $Q\equiv0$, and (\ref{55}) reduces to the formula given by Rosen. For the second example in the previous section, the tensor $P$ is covariantly constant with respect to (the symmetric connection) $\nabla$ and $P^{a}_{bc}l^{b}=0$. This establishes the equality of the Riemann tensors stated there.

\section*{References}
   
\end{document}